\documentstyle{mn}
\input psfig

\newif\ifAMStwofonts
\AMStwofontstrue

\def\apj#1{{ApJ,} {#1}}

\ifoldfss
  \ifCUPmtlplainloaded \else
    \NewTextAlphabet{textbfit} {cmbxti10} {}
    \NewTextAlphabet{textbfss} {cmssbx10} {}
    \NewMathAlphabet{mathbfit} {cmbxti10} {} 
    \NewMathAlphabet{mathbfss} {cmssbx10} {} 
  \fi
  \ifAMStwofonts
    \ifCUPmtlplainloaded \else
      \NewSymbolFont{upmath} {eurm10}
      \NewSymbolFont{AMSa} {msam10}
      \NewMathSymbol{\upi}     {0}{upmath}{19}
      \NewMathSymbol{\umu}     {0}{upmath}{16}
      \NewMathSymbol{\upartial}{0}{upmath}{40}
      \NewMathSymbol{\leqslant}{3}{AMSa}{36}
      \NewMathSymbol{\geqslant}{3}{AMSa}{3E}

      \let\leq=\leqslant 
       
    \fi
  \fi
\fi 

\ifnfssone
  \newmathalphabet{\mathit}
  \addtoversion{normal}{\mathit}{cmr}{m}{it}
  \addtoversion{bold}{\mathit}{cmr}{bx}{it}
  \newmathalphabet{\mathbfit} 
  \addtoversion{normal}{\mathbfit}{cmr}{bx}{it}
  \addtoversion{bold}{\mathbfit}{cmr}{bx}{it}
  \newmathalphabet{\mathbfss} 
  \addtoversion{normal}{\mathbfss}{cmss}{bx}{n}
  \addtoversion{bold}{\mathbfss}{cmss}{bx}{n}
  \ifAMStwofonts
    \ifCUPmtlplainloaded \else
      \UseAMStwoboldmath
      \makeatletter
      \new@mathgroup\upmath@group
      \define@mathgroup\mv@normal\upmath@group{eur}{m}{n}
      \define@mathgroup\mv@bold\upmath@group{eur}{b}{n}
      \edef\UPM{\hexnumber\upmath@group}
      \new@mathgroup\amsa@group
      \define@mathgroup\mv@normal\amsa@group{msa}{m}{n}
      \define@mathgroup\mv@bold\amsa@group{msa}{m}{n}
      \edef\AMSa{\hexnumber\amsa@group}
      \makeatother
      \mathchardef\upi="0\UPM19
      \mathchardef\umu="0\UPM16
      \mathchardef\upartial="0\UPM40
      \mathchardef\leqslant="3\AMSa36
      \mathchardef\geqslant="3\AMSa3E

      \let\leq=\leqslant 

    \fi
  \fi
\fi 

\ifnfsstwo
  \DeclareMathAlphabet{\mathbfit}{OT1}{cmr}{bx}{it}
  \SetMathAlphabet\mathbfit{bold}{OT1}{cmr}{bx}{it}
  \DeclareMathAlphabet{\mathbfss}{OT1}{cmss}{bx}{n}
  \SetMathAlphabet\mathbfss{bold}{OT1}{cmss}{bx}{n}
  \ifAMStwofonts
    \ifCUPmtlplainloaded \else
      \DeclareSymbolFont{UPM}{U}{eur}{m}{n}
      \SetSymbolFont{UPM}{bold}{U}{eur}{b}{n}
      \DeclareSymbolFont{AMSa}{U}{msa}{m}{n}
      \DeclareMathSymbol{\upi}{0}{UPM}{"19}
      \DeclareMathSymbol{\umu}{0}{UPM}{"16}
      \DeclareMathSymbol{\upartial}{0}{UPM}{"40}
      \DeclareMathSymbol{\leqslant}{3}{AMSa}{"36}
      \DeclareMathSymbol{\geqslant}{3}{AMSa}{"3E}

      \let\leq=\leqslant 

    \fi
  \fi
\fi 

\ifCUPmtlplainloaded \else
  \ifAMStwofonts \else 
    \def\upi{\pi}
    \def\umu{\mu}
    \def\upartial{\partial}
  \fi
\fi

\title[Thermal instability]
{Statistical theory of thermal instability}
\author[A. F. Illarionov and I. V. Igumenshchev]
       {Andrei F. Illarionov$^{1}$
       and Igor V. Igumenshchev$^{2,3}$ \\
        $^1$P. N. Lebedev Physical Institute, 84/32 Profsoyuznaya Street,
           Moscow, 117810, Russia \\
        $^2$Department of Astronomy \& Astrophysics, G{\"o}teborg
           University and Chalmers University of Technology,
           412 96 G{\"o}teborg, Sweden \\
	$^3$Institute of Astronomy, 48 Pyatnitskaya Street, 
 	   Moscow, 109117, Russia \\}

\date{Accepted 1997 September 00.
      Received 1997 September 00;
      in original form 1997 September 00}

\pagerange{\pageref{firstpage}--\pageref{lastpage}}
\pubyear{1997}

\begin{document}

\maketitle

\label{firstpage}

\begin{abstract}
A new statistical approach is presented
to study the process of the thermal instability
of optically thin unmagnetized
plasma. In the frame of this approach the time
evolution of mass distribution function over temperature
$\varphi(T)$ is calculated. Function $\varphi(T)$ characterizes
the statistical properties of the multiphase medium of arbitrarily
spaced {\it three-dimensional} structure of arbitrary (small or large)
temperature
perturbations.
We construct our theory under
the isobarical condition ($P=const$ over space), which is satisfied in
the
short wavelength limit of the perturbations.
The developed theory is illustrated in the case of thermal instability
of a slowly expanding interstellar cloud (smooth scenario).
Numerical
solutions of equations of the statistical theory are constructed and
compared
with hydrodynamical solutions.
The results of both approaches are identical in the short wavelength
range when the isobarity condition is satisfied.
Also the limits of applicability of the statistical theory are estimated.
The possible evolution of initial spectrum of perturbations is discussed.
The proposed theory and numerical models can be relevant to the
formation of the two-phases medium in the $\sim 1 pc$ region around quasars.
Then small warm ($T\simeq 10^4 K$)
clouds are formed as the result of thermal
instability in
an expanded gas fragment, which is a product of either the star-star or
star-accretion disc collision.
\end{abstract}


\section{Introduction}

The interstellar and intergalactic optically thin
        \footnote[0]{$^\star$E-mail: illarion@dpc.asc.rssi.ru (AFI);
         ivi@fy.chalmers.se (IVI)}
plasma may be in a variety of thermal phases, depending
on local heating and cooling processes and past history.
The equilibrium of the thermal phases in the case of thermal and
ionization balance and in the case of constant pressure of the
plasma have been
widely discussed (for references see, e.g., Lepp et al. 1985;
McKee \& Begelman 1990). More recent investigations are concentrated on
the process of two-phases medium formation
(Aranson, Meerson \& Sasorov 1993)
and dynamics of this medium (Elphick, Regev \& Shaviv, 1992; Aharonson, Regev
\& Shaviv 1994).

We focus our attention on the dynamics of the thermal
phases generation and separation of mass over temperatures due to the
the thermal instability process.
We are interested in the evolution of the most rapidly growing modes
which have time scales (and corresponding spacial scales) comparable
with the characteristic cooling time $t_c$.
The growth of small amplitude
perturbations of temperature and of density of the optically thin medium
under the action of external heating and radiative cooling
was studied by Weymann (1960) and Field (1965) in the linear approximation.
The nonlinear regime
is traditionally investigated by solving
the exact or reduced
hydrodynamical equations for the density, velocity and temperature
perturbations of gas in space (see, e.g., Aranson et al. 1993).
But, the detailed evolution models
in these approaches can be only constructed by assuming
the one-dimensional (1D) distributions.
The attempts to solve this problem in
two- and three-dimensions meet significant technical difficulties.

In the present paper we propose a new statistical approach to
discribe the dynamics of the thermal instability and construct
the theory for any (linear or nonlinear)
stages in isobaric approximation. In the frame of this approach we
investigate the time evolution of the mass distribution function
over temperature $\varphi(T)$ [see equation (3.3)], which characterizes
the statistical
properties of the multiphase medium.
Such a medium may contain arbitrarily spaced three-dimensional
perturbations of short scales. We assume the constant gas pressure ($P=const$)
over volume even
large contrasts of density $\rho$ and
temperature $T$ exist.
But, at the same time, the pressure is a {\it time dependent} value.
So, at any moment $t$ the density and temperature of matter
meet the relation $\rho\propto P/T$.

We show in Section~3  that in the case of the isobaric and nonuniform medium
the temperature growth rate $\dot{T}$ of the
matter with temperature $T$ [see equation (3.12)] is
\begin{equation}
   \dot{T}\propto -\left[ {1\over \gamma}{\cal L}+ (1-{1\over \gamma})
   \overline{\cal L}{T\over\overline{T}}\right]. 
\end{equation}
Here ${\cal L}={\cal L}(\rho,T)={\cal L}(P/T,T) $
is the local cooling-heating rate of the gas,
$ \overline{\cal L}$ and $\overline{T}$ are the mass average
values of ${\cal L}$ and $T$, respectively [see equation (3.11)], and
$\gamma$ is the gas adiabatic index.
In the case of the uniform medium the equation (1.1) reduces to the form
$\dot{T}\propto -{\cal L}$.
One can see that the growth rate (1.1) depends on the local $T$ only.
Other quantities are common for any parts of
the medium. They are  the average
over mass parameters, but are not independent variables.
In this case $\dot{T}=\dot{T}(T)$ the  time evolution equation
of the function $\varphi(T)$
[see equation (3.15)]
is of the Liouville type:
\begin{equation}
   {\partial\varphi\over\partial t}+{\partial\over\partial T}
   (\dot{T}\varphi)=0.
\end{equation}
These coupled equations (1.1) and (1.2) determine the evolution of
the medium on any (linear or nonlinear) stages.

We illustrate our statistical approach in the case
of a slowly expanding interstellar cloud when
the medium smoothly evolves from the thermally stable stage
to the unstable stage.
This evolution occurs through the critical point of marginal stability.
We keep in mind the evolution of the initially `warm' ($T\simeq 10^4$K)
gas
because it can be in a stable equilibrium in a very wide range of the
external
heating and ionization parameters.
The initial stages of the thermal instability
follow the linear theory of Weymann (1960) and Field (1965).
On the subsequent nonlinear stages, when the linear theory is not applicable,
we compare the numerical solutions of the equations (1.1) and
(1.2), and 1D hydrodynamical solutions.
The results of both statistical and hydrodynamical
approaches are identical in the
isobaric limit, when the pressure is almost constant over the medium.
We also estimate the limits of applicability of the statistical approach.

The hydrodynamical approach is discussed in Section~2.
In Section~3 the statistical theory of thermal instability is constructed
in the isobaric approximation.
The smooth scenario of the thermal instability process is considered in
Section~4.
In Section~5 we discuss the numerical solutions of the equations
of the statictical approach and
hydrodynamical equations, and
compare the results of both approaches.
In Section~6 we discuss the obtained results and the limits of
the constructed theory.
In Appendix~A we describe the numerical procedure of solution of the
statistical equations.
In Appendix~B we present the results of the linear theory in the
considered case of the smooth scenario.

\section[]{HYDRODYNAMICAL APPROACH}

\setcounter{equation}{0}

The dynamics of a thermally unstable plasma can be described by
the hydrodynamical
equations for ideal gas with the additional cooling-heating term in the
energy equation.
\begin{equation}
   {d\rho\over dt}+\rho\nabla\vec{v}=0,
\end{equation}
\begin{equation}
   \rho{d\vec{v}\over dt}+\nabla P=0,
\end{equation}
\begin{equation}
   {1\over\gamma-1}{dP\over dt}+{\gamma\over\gamma-1}P\nabla\vec{v}+
   \rho{\cal L}(\rho,T)=0,
\end{equation}
\begin{equation}
   P={R\over\mu}\rho T.
\end{equation}
Here $\rho$, $P$, $T$, $\vec{v}$ are the density, the pressure,
the temperature, and the velocity of gas, respectively, $R$ is the
gas constant, $\mu$ is the mean molecular weight, and $\gamma$ is the
adiabatic index. For simplicity, we do not take into account the
influence of the ionization processes on thermal balance, and assume
$\gamma$ and $\mu$ to be constant. We assume fully ionized hydrogen plasma,
setting $\gamma=5/3$ and $\mu=1/2$ for numerical estimations.
The cooling-heating rate ${\cal L}$ is defined
as the rate of energy losses minus the heating rate per unit mass.
We assume that ${\cal L}$ depends on the local density $\rho$ and
temperature $T$ of gas. In the simplest case of low density and
optically thin plasma (Spitzer 1962) this rate is
\begin{equation} 
  {\cal L}(\rho,T)=\rho\Lambda(T)-H, 
\end{equation}
where $\Lambda(T)$ and $H$ are cooling and heating rates, respectively.
The heating rate of the plasma can be determined by different mechanisms
(e.g., photoabsorbtion, Compton-effect, heating due to cosmic rays
and nuclear decays). For generality, we do not specify here these
mechanisms and use a constant value for the heating rate in our models.
In the simplest case of a stationary medium of
temperature $T_0$ and
density $\rho_0$ the criterion of marginal stablibility
takes the form (Weymann 1960, Field 1965)
\begin{equation} 
   \left.{d\over dT}\left({\Lambda\over T}\right)\right|_{T_0}=0,\;\;\;\;
   \rho_0={H\over\Lambda(T_0)}.
\end{equation}
In the stable region the derivative is positive, in the unstable region
it is negative. The cooling function $\Lambda(T)$
was calculated in
a number of papers
(e.g., Cox \& Tucker 1969; Raymond, Cox \& Smith 1976)
for the different conditions in the interstellar plasma.
For typical conditions of thermal interstellar plasma
the marginal temperature $T_0$ is about $10^4 K$.
The function $\Lambda(T)/T$ has a maximum at $T=T_0$.
At $T<T_0$ the plasma
is thermally stable, and at $T>T_0$ the plasma is unstable.
In the presence of external ionizing (UV) radiation the function
$\Lambda(T)$ changes significantly
(e.g., Krolik, McKee \& Tarter 1981).

In equation (2.3) we omit the thermal conductivity term
$\nabla(\kappa\nabla T)$, where $\kappa$ is the
thermal conductivity coefficient.
The thermal conductivity process stabilizes the thermal instability
at the shortest wavelengths $\lambda\la 2\pi(\kappa\mu/nR\rho)^{1/2}$,
where $n$ is the growth rate of instability
(see the discussion in Section~6).
The modes of larger wavelength can be developed till the nonlinear stages.
For these modes
the thermal conductivity process is important at the nonlinear stage,
when the medium becomes very nonuniform and is divided into
cold and hot phases.
The thermal conductivity effectively stabilizes the further temperature
growth
of the hot phase and can produce the evaporation of the cold phase.
In a subsequent paper, we will analyze
the influence of thermal conductivity at the nonlinear stage of the
thermal
instability process.

\section[]{STATISTICAL APPROACH}

\setcounter{equation}{0}

We consider the cloud of plasma of mass $M$ and of volume $V$.
The equations (2.1)-(2.4) give us a precise description of the
thermal and dynamical evolution of this matter. However,
the solution of the problem in general case
meets significant technical difficulties.
We simplify the analysis proposing a new statistical approach
which is constructed under isobaric approximation.
The isobaric condition corresponds
to a short wavelength limit
\begin{equation}
   \lambda \la \lambda_P = 2\pi c_s t_c,
\end{equation}
when the pressure balance time
$\sim \lambda/c_s$ is shorter than the characteristic thermal time
\begin{equation}
   t_c={c_s^2\over H}.
\end{equation}
Here $c_s=(\gamma P/\rho)^{1/2}$ is the adiabatic sonic speed.
The hydrodynamical calculations confirm (see Section 5) that
in the wavelength range (3.1)
the isobaric approximation is satisfied.
The calculations also set the limits
of applicability of the statistical approach.

We build our theory using
the temperature-dependent distribution of matter
instead of the space-dependent one.
We sum the  masses of equal
temperature and compose the distribution function $\varphi(T)$
of mass over temperature.
Namely, the mass of gas $\Delta m$ having the  temperature
in the range from $T$ to $T+\Delta T$ is
\begin{equation}
   \Delta m=M\varphi(T)\Delta T, 
\end{equation}
where the distribution function $\varphi(T)$ is
normalized to unity
\begin{equation}
   \int_0^\infty\varphi(T)dT=1.
\end{equation}

Now we show that the temperature growth rate $\dot{T}$ depends only on
local $T$ and the global (average) parameters of the medium.
First we prove that for isobaric perturbations the pressure of the gas
\begin{equation}
   P={R\over\mu}{M\over V}\overline{T}
\end{equation}
is determined by the mass average temperature
\begin{equation} 
  \overline{T}=\int_0^\infty T\varphi(T)dT ,
\end{equation}
and the average density of the cloud
$\langle\rho\rangle=M/V$.
Indeed, using equations (2.4) and (3.3) we can write
\begin{equation}
   V=\int_M{dm\over\rho}=\int_0^\infty {M\varphi(T)dT\over\rho}=
  {R\over\mu}M \int_0^\infty {T\varphi(T)dT\over P}.
\end{equation}
In the isobaric case $P=const$ the equation (3.5) follows from (3.7) directly.

\noindent
Second, from equations (2.1) and (2.4) we have for the velocity divergency
\begin{equation}
   \nabla\vec{v}= -{\dot{\rho}\over\rho}= -{\dot{P}\over P}
   +{\dot{T}\over T}.
\end{equation}
Substituting $\nabla\vec{v}$ into equation (2.3) we obtain
the relation between temperature and pressure growth rates
\begin{equation}
   \dot{T}={\gamma-1 \over\gamma}\left({\dot{P}\over P}T-
   {\mu\over R}{\cal L}\right). 
\end{equation}
Third, taking the average in (3.9)
we set $\overline{T}=PV\mu/MR$ [see equation (3.5)]
instead of $T$ in the right side
and correspondingly $\dot{(PV)}\mu/MR$ instead of $\dot{T}$ in the
left side, and find for the pressure derivative
\begin{equation}
   \dot{P}=-(\gamma-1){M\over V}\overline{\cal L}-
   \gamma{\dot{V}\over V}P.
\end{equation}
Here and below the mass average of any function $A(T)$ is defined as
\begin{equation}
   \overline{A}=\int_0^\infty \varphi(T)A(T)dT. 
\end{equation}
Finally, substituting $\dot{P}$ from (3.10) into (3.9) we  have the
temperature growth rate of the matter with
temperature $T$:
\begin{equation}
   \dot{T}=-{\gamma-1\over\gamma}{\mu\over R}\left[(\gamma-1)
   \overline{\cal L}{T\over\overline{T}}+{\cal L}\right]-
  (\gamma-1){\dot{V}\over V}T.
\end{equation}
This equation follows from the mass (2.1) and energy (2.3)
conservation equations. Instead of the Eulear equation (2.2)
we use the isobaric condition $P=C=const$.
Hydrodynamical calculations give us that $P=C+O(\lambda^2)$
(see Appendix~B), but we ignore here the additional
term $O(\lambda^2)$ in the $\lambda\la\lambda_P$ limit.

Note, that small velocities on small scales provide a
scale-independent value of the
velocity divergency, $\nabla\vec{v}\sim v_\lambda/\lambda$.
One can substitute (3.10) and (3.12) into (3.8) and obtain explicitly
\begin{equation}
   \nabla\vec{v}={\gamma-1\over\gamma}{\mu\over R}\left(
   {\overline{\cal L}\over\overline{T}}-{{\cal L}\over T}\right)+
   {\dot{V}\over V}, 
\end{equation}
which is determined only by the local temperature $T$.

In the isobaric case, according to (2.4) and (3.5) we have for density
$ \rho=\langle\rho\rangle\overline{T}/T$.
It has to be set in the function
${\cal L}(\rho,T)$ everywhere.
For example, the rate ${\cal L}$, determined by formula (2.5),
and its average value $\overline{\cal L}$, both of which
we need to put into equations (3.12) and (3.13),
have forms
\begin{equation}
   {\cal L}=\langle\rho\rangle{\Lambda(T)\over T}\overline{T}-H,\;\;\;\;\;\; 
   \overline{\cal L}=\langle\rho\rangle
   \overline{\left({\Lambda\over T}\right)}\overline{T}-H.
\end{equation}
After that we have all of the parameters
(${\cal L}$, $\overline{\cal L}$, $\overline{T}$) to substitute to
formula (3.12) for the rate $\dot{T}$. Also we have formulae for
the pressure (3.5) and the velocity divergency
(3.13) of the medium.

The time evolution of the mass distribution $\varphi(T)$ we
describe by the Liouville type equation
\begin{equation}
   {\partial\varphi\over\partial t}+{\partial\over\partial T}
   (\dot{T}\varphi)=0,
\end{equation}
where the temperature growth rate $\dot{T}$ is given by formula (3.12).
The equation (3.15) is
nonlinear because the term $\dot{T}$ depends on
the function $\varphi(T)$ through the average values
$\overline{T}$ and $\overline{\cal L}$.
If we specify the initial conditions at the moment $t_i$,
the cooling-heating function ${\cal L}(\rho,T)$
and the law of  expansion of the medium $V(t)$, then
the equations (3.12)
and (3.15) describe the thermal evolution of medium for
$t>t_i$.
The  initial conditions are the
distribution function $\varphi(T,t_i)$ and
mean density $\langle\rho\rangle_i$.
To solve the equations (3.12) and (3.15) we use
a numerical technique (see Appendix A)
in the general case of finite amplitude perturbations.

The evolution equation
for the relative temperature perturbation $\delta T/\overline{T}$,
where $\delta T=T-\overline{T}$,
can be constructed
using equation (3.12)
for $\dot{T}$ and its average version for $\dot{\overline{T}}$.
One can get
\[ {d\over dt}\left({\delta T\over\overline{T}}\right)=
   {\gamma-1\over\gamma}{\mu\over R\overline{T}}\left(
    \overline{\cal L}{T\over\overline{T}}-{\cal L}
   \right). \]
Under linear approximation we can write
$ {\cal L}(T)={\cal L}(\overline{T})+
  \left(\partial{\cal L}/\partial T\right)_P\delta{T}$
and find
$\overline{\cal L}={\cal L}(\overline{T})$, because of
$\overline{\delta T}=0$. In this case the equation
for the relative temperature perturbation becomes
\begin{equation}
   {d\over dt}\left({\delta T\over\overline{T}}\right)=
   {\gamma-1\over\gamma}{\mu\over R\overline{T}}\left({\cal L}-
  \left({\partial{\cal L}\over\partial T}\right)_P \overline{T}
   \right){\delta T\over\overline{T}}. 
\end{equation}
Here the values of ${\cal L}$ and
$\left({\partial{\cal L}/\partial T}\right)_P$
are taken at $T=\overline{T}$.
For example, in case of the
static uniform medium at the thermal equilibrium ${\cal L}(\overline{T})=0$
we find the growth rate
\begin{equation}
    n={\gamma-1\over\gamma}{\mu\over R\overline{T}}\left(
    \langle\rho\rangle
    \left({\partial{\cal L}\over\partial \rho}\right)_T-
    \overline{T}
    \left({\partial{\cal L}\over\partial T}\right)_\rho
    \right),
\end{equation}
which coincides with the growth rate found by Weymann (1960) and
Field (1965).
For the cooling-heating function (2.5) it is reduced to
$$ \!\!\,\, n={\gamma-1 \over t_c}\left(1-{T\over\Lambda}
   {d\Lambda \over dT} \right),
    ~~~\qquad\qquad\qquad\qquad\qquad~~\eqno(3.17')$$
where $t_c$ is given by (3.2). From equation $(3.17')$ one can obtain the
criterion (2.6).

Finally we note the following:

\noindent
1. One can construct uniquely the distribution function $\varphi(T)$ from any
$T(\vec{r})$
summing the correspondent masses of volumes
$\Delta V_i$ with temperatures
in range from $T$ to $T+\Delta T$,
\[ \varphi(T)\Delta T={\rho\sum_i \Delta V_i\over M}.\]
For instance, consider a one dimensional case and the cosine profile
of temperature,
$T(x)=T_0-\delta T_0\cos{(2\pi x/\lambda)}$.
Then we can obtain inverse dependence $x(T)$ for the interval $0<x<\lambda/2$
of monotonicity of $T(x)$ and find the distribution function
\begin{eqnarray}
   \varphi(T) & \!\!\!\! = & \!\!\!\! {2\over\lambda}{\overline{T}\over T}
   {dx(T)\over dT}={1\over\pi}{\overline{T}\over T}
   {1\over\sqrt{\delta T_0^2-(T-T_0)^2}}, \nonumber \\
              &      & \qquad\qquad T_0-\delta T_0<T<T_0+\delta T_0, 
\end{eqnarray}
where the mass average
$\overline{T}=\sqrt{T_0^2-\delta T_0^2}$, and $\delta T_0<T_0$.
The dispersion $\sigma_T^2=\overline{(T-\overline{T})^2}/\overline{T}^2
\approx{1\over 2}(\delta T_0/T_0)^2 $ at $\delta T_0 \ll T_0$.
The function (3.18)
is independent of the space period $\lambda$ and has two integrable
infinities at $T_{min}=T_0-\delta T_0$ and
$T_{max}=T_0+\delta T_0$ as shown in Fig.~1.
\begin{figure}
 \hbox{\psfig{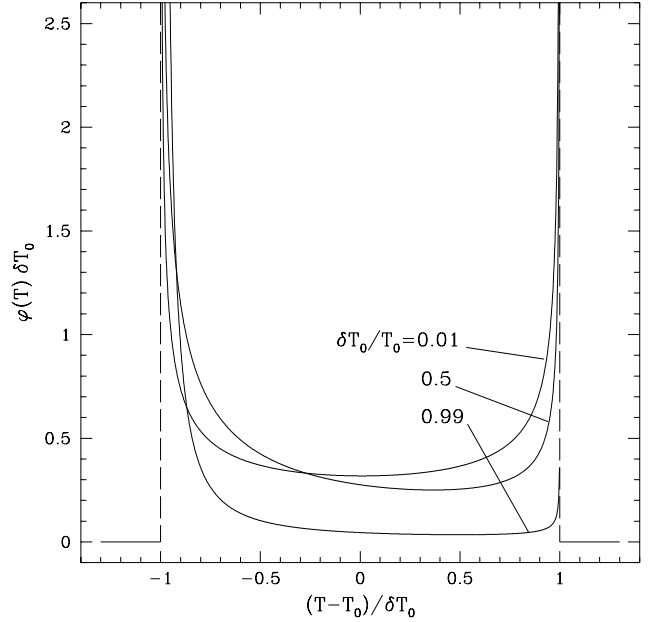}}
 \caption{
The distribution function $\varphi(T)$ in the case of the cosine profile
$T(x)=T_0+\delta T_0 \cos(2\pi x/\lambda)$
for different values of relative amplitude $\delta T_0/T_0=0.01,\;0.5,\;0.99$.
}
\end{figure}

\vskip 0.3truecm
\noindent
2. The reconstruction of the space distribution $T(\vec{r})$,
$\rho(\vec{r})$ and velocity field $\vec{v}(\vec{r})$
from a given $\varphi(T)$ is an incorrect procedure.
The recovery can be uniquely done in the one dimensional case
assuming additional restrictions on the temperature profile $T(x)$.
The temperature must be a periodical, monotonical for one half period
($0<x<\lambda/2$) and odd symmetrical (relative $x=0$) function.
Keeping this restrictions one can find the implicit dependence $T(x)$:
\begin{equation}
   2{x\over\lambda}={1\over\overline{T}}\int_{0}^{T(x)}
   T' \varphi(T')dT'.
\end{equation}
Then for the density distribution we have
\begin{equation}
  \rho(x)=\langle\rho\rangle\overline{T}/T(x). 
\end{equation}
The velocity profile can be obtained by integrating equation (3.13):
\[   v(x) =  \tilde{v}(x)+{\dot{V}\over V}x  = \]
\begin{equation}
   {\gamma-1\over\gamma}{\mu\over R\overline{T}}
   {\lambda\over 2}\int_0^{T(x)}\left(
   {\overline{\cal L}\over\overline{T}}T'-{\cal L}(T')\right)
   \varphi(T')dT'
   +{\dot{V}\over V}x.
\end{equation}
The amplitude of the perturbed part
$\tilde{v}(x)$ of $v(x)$ is proportional to
the space period $\lambda$, and
$\tilde{v}(0)=\tilde{v}(\lambda/2)=0$.

\section[]{SMOOTH SCENARIO}

\setcounter{equation}{0}

We illustrate our statistical theory of thermal instability
in the case of inertialy expanding uniform  plasma cloud.
The density of the cloud
decreases with time, say,
\begin{equation}
   \rho_u(t)\propto{1\over t^3}.
\end{equation}
Initially the plasma is assumed to be quite dense and cold.
The temperature  $T_u$ of the uniform plasma is less than the marginal
value $T_0$, and the plasma is thermally and mechanically stable
[see condition (2.6) and subsequent discussion].
In the case of a constant heating rate $H$ the temperature $T_u$
increases up to the marginal temperature $T_0$. It happens at some moment
$t_0$.
At time $t>t_0$ the plasma is unstable.

Numerical models show (see Section~5.1) that the fragmentation of cloud
is only possible in the
case of a `slowly' expanding cloud for which the characteristic thermal
time $t_c$ [equation (3.2)] is much shorter than
expanding time $t_0$. In this case $t_c\ll t_0$
the thermal (quasi) equilibrium of uniform
matter is established, ${\cal L}=0$.
Then the evolution of temperature $T_u(t)$ follows from
the equation
\begin{equation}
   \Lambda(T_u)={H\over \rho_u}=\Lambda(T_0)\left(t\over t_0 \right)^3.
\end{equation}
Near the marginal point $T_0$ (where $ \Lambda= T d\Lambda/dT$)
one can find
$$  \!\!\,\, T_u(t)=T_0(1+3{\delta t\over t_0}),
    \qquad\qquad\qquad\qquad\qquad\qquad\quad~~\eqno(4.2')$$
where $\delta t=t-t_0\ll t_0$.

In the thermally unstable plasma the fastest growth rate
has a condensation mode which corresponds to the conditions
$\lambda\la\lambda_P$ and $P\approx const$.
In the linear regime the amplitude $A$ of this mode follows the equation
$\dot{A}=n A$. At $\delta t\ll t_0$
the growth rate $n$ is small and
linearly proportional to $\delta t$:
\begin{equation}
   n(t)=3b{\gamma-1\over t_c}{\delta t\over t_0},
\end{equation}
where $b=-T_0^2\Lambda ''(T_0)/\Lambda(T_0)$ is the positive quantity.
The formula (4.3) follows from equation $(3.17')$, where
the term in the brackets can be linearly approximated
$1-T\Lambda '/\Lambda=-T_0(T-T_0)\Lambda ''/\Lambda$
at the limit $T-T_0\ll T_0$.
In addition the time dependency of temperature $(4.2')$ is used.
Here and below we define the parameter $t_c=c_s^2(T_0)/H$ at the moment
$t=t_0$.
On the linear stage ( Weymann 1960, Field 1965, see
also Appendix B) at
$0<\delta t<\tau$, the amplitude increases as
\begin{equation}
   A(t)=A_0\exp{N(t)},\;\;\;\;\;
  N(t)=\int_{t_0}^t
  n dt=
  {3\over 2}b{\gamma-1\over t_c}
   {\delta t^2\over t_0},
\end{equation}
and we must study the progressively inhomogeneous medium.
Equation (4.1) is then correct for the average density
$\langle\rho\rangle$ only.
The linear regime is limited by the time interval $0<\delta t <\tau$, where
the end time
\begin{equation}
  \tau=\sqrt{{2 t_0t_c\ln{(1/A_0)} \over 3b(\gamma-1)}}.
\end{equation}
At $\delta t=\tau$ the integral $N=\ln {(1/A_0)}$ and the amplitude $A=1$.
At $t>t_0+\tau$ the perturbations are large and grow nonlinearly.
In case of the small parameter $t_c/t_0\ll 1$, the time interval
of the linear stage is also small:
$\tau/t_0\approx\sqrt{\ln{(1/A_0)} t_c/t_0}\ll 1$.

The scenario of the expanding cloud is able to follow the natural smooth
evolution of the medium from the stable ($t<t_0$, $T<T_0$) to the unstable
($t>t_0$, $T>T_0$) stages through the critical point ($T=T_0$).
Other scenarii of a smooth transition through the critical point $T=T_0$
can also be constructed. For example, we can take the stationary medium
($\rho_u=\rho_0=const$) heated at a slowly increasing rate and
obtain similar results assuming $t_c\ll t_0$ and setting
$H(t)=H_0(t/t_0)^3$,
$H_0=\rho_0 \Lambda(T_0)$.

\section[]{NUMERICAL RESULTS}

\setcounter{equation}{0}

We use the expanding cloud scenario to numerically study
the thermal instability
process in the framework of statistical isobaric theory.
We find numerically the time evolution of the distribution
function $\varphi (T,t)$ and the evolution of corresponding
characteristic parameters of this distribution.
We also find the final mass fraction of cold and hot phases as a
function of the parameters of the problem.
To test the results of the statistical approach
we compare them with the results of hydrodynamical calculations
in the following way. We use the obtained $\varphi (T,t)$ to construct
the $T(x)$, $\rho (x)$ and $v(x)$ profiles. To do this we use the
reconstruction procedure, which is discussed at the end of Section~3
[see equations (3.19), (3.20) and (3.21)].
After this we compare the constructed profiles with the profiles
resulting from the 1D hydrodynamical calculations under the identical
parameters of the problem and the similar initial conditions.
In all models we fix the
heating rate $H=const$ and set $\gamma=5/3, \mu=1/2$.
The expansion law (4.1) for the average density $\langle\rho\rangle$
of the plasma is assumed in the calculations.

We use the model curve of the cooling function $\Lambda(T)$, which mimics
the base properties of the cooling processes in the interstellar plasma,
\begin{equation}
   \Lambda(T) = \Lambda_0{2(T/T_0)^2\over 1+(T/T_0)^2}\theta(T),
\end{equation}
where the function $\theta(T)$ is
\[ \theta(T)= \left\{ \begin{array}{ll}
                        1 , & \mbox{if $T/T_0>c$,} \\
       \exp[d(1/c-T_0/T)] , & \mbox{otherwise.}
		      \end{array}  \right. \]
Here we assume
$c=0.9$ and  $d=10.0$.
The exponential drop of $\theta(T)$ in the low temperature region
$T< c T_0$ mimics the strong fall of $\Lambda (T)$.
The cooling curve (5.1) has
the critical point $T=T_0=10^4 K$,
where the criterion of marginal stability (2.6)
is fulfilled.
The instability growth rate coefficient [see equation (4.3)] $b=1$ in this case.
We note, that the cooling curve (5.1) does not allow a stable
equilibrium for a two-phases medium.

We start our calculations at moment $t=t_0$.
The initial relative perturbations of temperature
$\Delta=(T_{max}-T_{min})/\overline{T}$ and the expansion parameter
$t_0/t_c=t_0 H/c_s^2(T_0)$ are free parameters of the models.
The hydrodynamical models, in addition, are characterized by
the initial space period $\lambda_0/\lambda_P$ of perturbations.
Here $\lambda_P=2\pi c_s^3(T_0)/H=2\pi c_s t_c$.

The cosine profile of temperature
$T(x)=T_0\left(1-{\Delta\over 2}\cos (2\pi x/\lambda_0)\right)$
is used as the initial perturbations in the hydrodynamical models.
In the statistical models the corresponding initial distribution
function $\varphi (T)$ is given by equation (3.18).
The  initial temperature dispersion $\sigma_T^2(t_0)=\Delta^2/8=A_0^2/2$.

\subsection[]{Statistical isobaric models}

The numerical method used to solve the equations (3.12) and (3.15)
of statistical approach
is briefly described in Appendix A.

\begin{figure}
 \hbox{\psfig{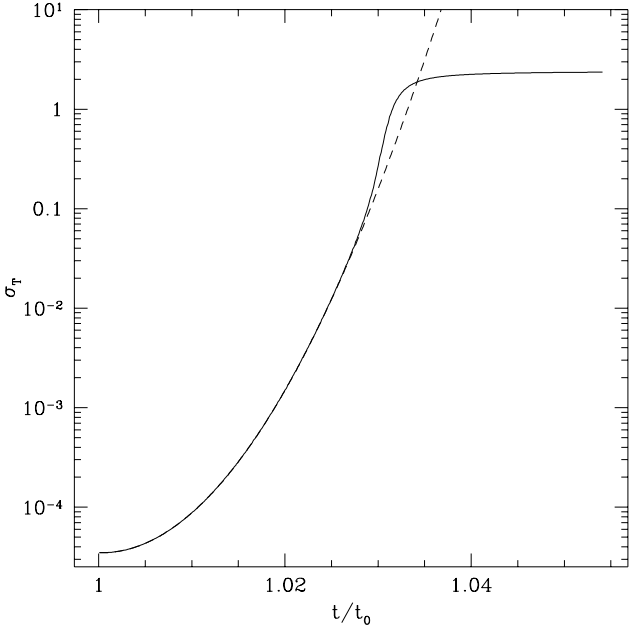}}
 \caption{
The evolution of the relative perturbation
of temperature $\sigma_T$ for the statistical isobaric S-model.
The solid line shows the evolution coming from nonlinear theory, and
the dashed line shows the result in the linear approximation.
}
\end{figure}

We discuss, as a characteristic example, the detailed evolution
of the S-model, with the initial parameters
$\Delta=10^{-4}$ and $t_0/t_c=10^4$.
The temperature dispersion
$\sigma_T^2=\overline{(T-\overline{T})^2}/\overline{T}^2$
increases with time at the $t>t_0$ stage. The evolution of $\sigma_T(t)$
is shown in Fig.~2.
Initially $\sigma_T(t_0)=\Delta/\sqrt{8}$.
The solid line corresponds to the nonlinear calculation, and
the dashed curve
was calculated under the linear approximation [equation (3.16)].
As it was shown in Section 4 [equation (4.4)] at the linear stage
$\sigma_T(t)\approx\sigma_T(t_0)\exp{[(t-t_0)^2/t_0 t_c]}$.
A significant deviation is observed between the linear and nonlinear
stages when the value of $\sigma_T$ is large, $\sigma_T\ga 0.1$.
The instability results in the formation of the two-phases medium
at the moment $t\approx t_0+\tau$, where the time interval $\tau$
is determined by equation (4.5).

\begin{figure*}
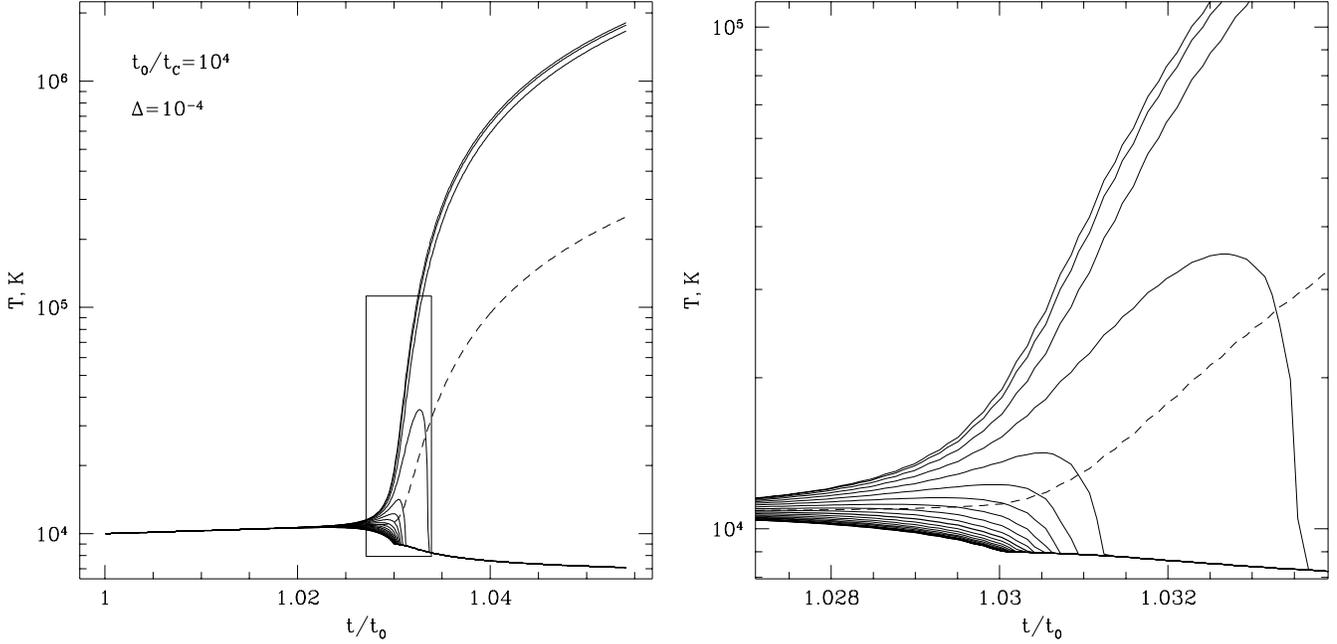

 \hbox{\psfig{file=tins_fig3a.epsi,height=8.5cm}
 \hbox to 2mm {} 
       \psfig{file=tins_fig3b.epsi,height=8.5cm}}
 \caption{
The evolution of temperatures of the fixed mass coordinates (solid lines)
and average temperature $\overline{T}$ (dashed line)
for the S-model.
The mass coordinates of neighbouring solid lines  differ by
$\Delta M/M=0.05$.
The upper and lower lines correspond to maximum and minimum temperature
of plasma.
The long time evolution is shown in the left-hand panel.
Detailed evolution at right represents the nonlinear stage
shown inside the box in the left-hand panel.
}
\end{figure*}

In Fig.~3 (left- and right-hand panels) we show the trajectories
of mass particles in (T,t)-plane.
Each line is the trajectory
of a fixed mass coordinate, and for each two neighbour lines
$\ell_1$ and $\ell_2$ the following condition is satisfied:
\[{\Delta M\over M}=\int^{T(\ell_2)}_{T(\ell_1)}\varphi(T')dT'=0.05, \]
for any time $t$.
The upper and lower lines correspond to maximum and
minimum temperatures of plasma. The dashed line shows
the evolution of average temperature $\overline{T}(t)$.
Note, that the time dependence of $\overline{T}(t)$ does not coincide
with the evolution of temperature of the uniform medium followed from
equation (4.2).
In the linear stage of instability the lines are spaced closely.
In the nonlinear stage,
which is approximately shown inside the box in the left-hand panel
and in more detail in the right-hand panel of Fig.~3,
the lines disperse rapidly.
One group of lines goes up to the high temperatures,
and another group goes down
to the low temperatures.
A small part of gas of intermediate temperatures shows nonmonotonic
behaviour in its temperature evolution.

At the following time $t>t_0+\tau$ the two-phase medium
with hot ($T_h$, $\rho_h$) and cold ($T_c$, $\rho_c$) phases
is formed. A sharp boundary separates
phases.
Numerical calculations show that the mass fraction of the cold
and hot phases tend
asymptotically  to final values.
In the case of the S-model the final fraction
of the cold mass is $\eta=0.86$.
The remainder $1-\eta=0.14$ is the final mass fraction of the hot matter.
The temperature dispersion of this medium is
\[ \sigma^2_T=\left({T_h\over\overline{T}}-1\right)^2(1-\eta)+
   \left(1-{T_c\over\overline{T}}\right)^2\eta. \]
In the limit $T_c\ll T_h$ one obtains a simple relation
$\sigma^2_T=\eta/(1-\eta)$. For the case $\eta=0.86$, we find
$\sigma_T=2.5$, which very well agrees with our numerical
result (see Fig.~2).

Consider now the evolution of the parameters of hot
and cold
phases at the stage $t-t_0\ll t_0$,
when the expansion can be neglected.
Assuming $T_h\gg T_c$ one can
substitute in equation (2.3) the density of hot gas
$\rho_h=\langle\rho\rangle_0 (1-\eta)$, where
$\langle\rho\rangle_0=H/\Lambda_0$, and obtain the equation:
\begin{equation}
   \dot{T_h}=(\gamma-1){\mu\over R}H\left({\Lambda(T_h)\over\Lambda_0}
   (\eta-1)+1\right).
\end{equation}
The ratio
$\Lambda(T_h)/\Lambda_0=2$ [see formula (5.1)]
and we obtain $\dot{T_h}\sim (2\eta-1)$.
The temperature of the hot phase increases
to infinity and a two-phase medium can be  formed
in the case when more than half
of the matter falls down to the cold phase, $\eta>1/2$.
Finally, at $t>t_0+\tau$, we have
\[ T_h\simeq (\gamma-1){\mu\over R}H(2\eta-1)(t-t_0-\tau). \]
The numerical calculations
confirm this estimation.
The temperature $T_c$ and density $\rho_c$
of cold phase are determined by the condition of thermal
equilibrium $H=\rho_c\Lambda(T_c)$ and the isobaric relation with
hot phase, which corresponds to the relation
$ \rho_c=\langle\rho\rangle_0\left[(1-\eta)T_h/T_c
   + \eta\right]$.

\begin{figure}
 \hbox{\psfig{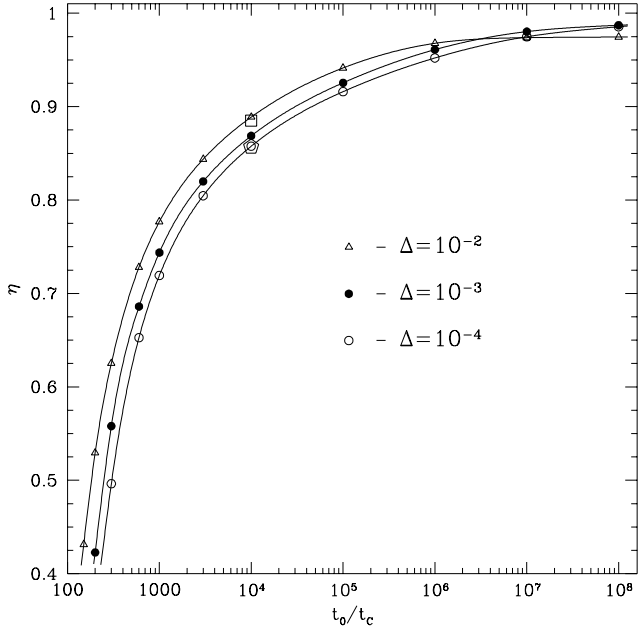}}
 \caption{
The final mass fraction of cold phase $\eta$ of the 
statistical isobaric models as a function of
the expansion parameter $t_0/t_c$
for the initial amplitudes $\Delta=10^{-2}$, $10^{-3}$ and $10^{-4}$
(deltas, full circles and open circles).
A pentagon and square show
the final mass fractions of the hydrodynamical H1 and H2-models,
respectively, which have the initial amplitude $\Delta=10^{-4}$.
The H1-model and isobaric S-model (at $t_0/t_c=10^4$ and $\Delta=10^{-4}$)
have almost equal final mass fractions.
}
\end{figure}

The final mass fraction of the phases is an important characteristic of
the models. We studied the dependence of the mass fraction $\eta$ on
the parameters
 $\Delta$ and $t_0/t_c$. Fig.~4  shows
$\eta$ as a function of $t_0/t_c$
for three values of the initial amplitude
$\Delta=10^{-2},\; 10^{-3}\; 10^{-4}$.
In the case of a slowly expanding limit
$t_0/t_c\ga 10^4$
the mass fraction $\eta$
is a slowly increasing function of $t_0/t_c$ and
weakly ($<5\%$) depends on  $\Delta$,
when $\eta\ga 0.85$. In case of a fast expansion
$t_0/t_c\la 10^3$, the mass fraction $\eta$ decreases, and its
dependence on $\Delta$ becomes significant.
The value of $\eta$ reaches the minimum $\approx 0.4$ at $t_0/t_c\approx 10^2$,
when the end time of the linear stage $\tau\simeq 0.3t_0$ [see equation (4.5)]
becomes equal to the expansion time 
$V/\dot{V}\simeq t_0/3$ [see equation ($4.1$)].
For a faster expanding medium, $t_0/t_c\la 10^2$,
the final two-phase structure is not developed.
In this case the expansion factor suppresses the thermal instability, and
the models show the initial growth of the dispersion $\sigma_T^2$
until some maximum and the following decrease.
The medium remains hot and uniform.

The reconstructed 1D space distributions of density
and perturbed velocity for one period $0<x<\lambda$ are
shown in Figs~6 and 7 as dashed lines.
The lines are presented at subsequent moments, and the period $\lambda$
increases with time as $\lambda(t)=\lambda_0 t/t_0$.
The used reconstruction procedure is discussed in Section~3.

\subsection[]{Hydrodynamical models}

To solve equations (2.1)--(2.4) in the one-dimensional case
we used an implicit hydrodynamical code. The code is based on
the Lagrangian fully conservative numerical scheme of
the second order accuracy
(Samarskij \& Popov 1980).

\begin{figure}
 \hbox{\psfig{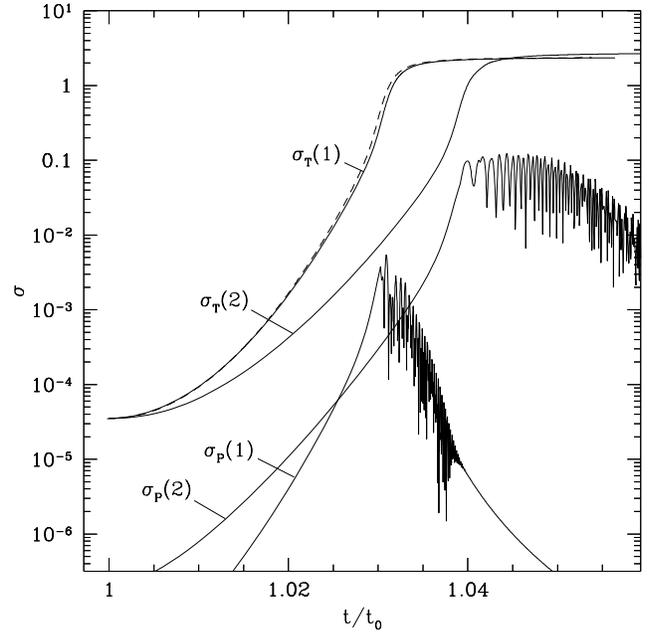}}
 \caption{
The evolution of relative perturbations of temperature $\sigma_T$ and pressure
$\sigma_P$ for H1-model and H2-model [solid lines, labeled
as (1) and (2), respectively]. The evolution of temperature perturbations
for the S-model (as in Fig. 2) is shown by a dashed line for
comparison.
}
\end{figure}

We discuss here two models, which differ in the initial
wave length $\lambda_0$ of perturbations, $\lambda_0=\lambda_P$ (H1-model) and
$\lambda_0=5 \lambda_P$ (H2-model).
They have the initial parameters $\Delta=10^{-4}$ and $t_0/t_c=10^4$,
which are identical to the S-model (see previous Section 5.1).
In addition, we assume that the initial perturbed velocities and pressure
variations equal to zero.
The computed characteristics of the S-model almost coincide with
those of the H1-model. However the H2-model shows significant differences
from both S and H1 models in the linear and nonlinear stages.
Fig.~5 shows the relative perturbations
(square root of the correspondent dispersion) of temperature $\sigma_T$ and
pressure $\sigma_P$ as functions of time for H1 and H2 models (solid lines).
In the same Fig.~5 the evolution of $\sigma_T$ for
the S-model is shown for comparison
(dashed line).
The differences between the curves $\sigma_T(t)$
for H1 and H2 models are explained by
the time delay. This time delay arises in the linear stage of evolution
due to differences of the growth rates
of different modes (see Appendix~B, Table~B1).
The perturbations for the shorter wave models grow more rapidly, and
the most rapid growth rate is observed for the isobaric S-model,
which corresponds to the limit $\lambda\rightarrow 0$.
The shape of $\sigma_T$ curves for H1 and H2 models
are very similar in the nonlinear stage
($\sigma_T>0.1$), but delayed in time.

To study the problem of nonlinear interaction of the modes with different
$\lambda_0$ we calculated the number of models where the initial conditions
were taken as a superposition of two modes, $\lambda_0=\lambda_P$
and $\lambda_0=5\lambda_P$. In the case of equal
initial amplitudes of the modes
the long wave mode was totally suppressed by the short
wave mode in the nonlinear phase. The long wave mode
suppresses the development of the short wave mode when the
ratio of the initial amplitudes $\ga 50$.
This result quantitatively well agrees with the estimations made in
the linear theory (Appendix~B).
Indeed, from Table~B1 one can find the difference of the exponential
factors of these two modes $\Delta N=9.56-6.0=3.56$.
It gives us the ratio of amplitudes $\exp(3.56)=35$.

The small value of relative pressure perturbation
$\sigma_P$ ($\la 10^{-2}$) is an indicator of the isobarity of the process.
Both H1 and H2 models have peak-like $\sigma_P(t)$ curves (see Fig.~5),
and the peaks are
located at the end of the linear evolution stage, when the most rapid
growth of temperature perturbations takes place.
In the case of the H1-model the maximum of $\sigma_P$
has a rather small value ${\sigma_P}_{max}\simeq 0.005$ and
even  at  the  nonlinear stage the process is almost isobarical.
This value of ${\sigma_P}_{max}$ can be compared with the correspondent
value followed from the linear theory, ${\sigma_P}_{max}\approx\alpha=0.006$
(see Appendix~B, Table~B1).
After the peak $\sigma_P$ decreases rapidly.
At this stage the sonic waves are generated
during the contraction of the cold phase.
These waves produce oscillations in $\sigma_P$ clearly seen in Fig.~5.
The intensity of the oscillations also rapidly decrease with time.
In the case of the H2-model the relative perturbations of pressure
reaches a rather high
value ${\sigma_P}_{max}\simeq 0.1$
(the linear theory gives 0.067, see Table~B1) and
slowly decreases in oscillation regime
after the maximum.

\begin{figure}
 \hbox{\psfig{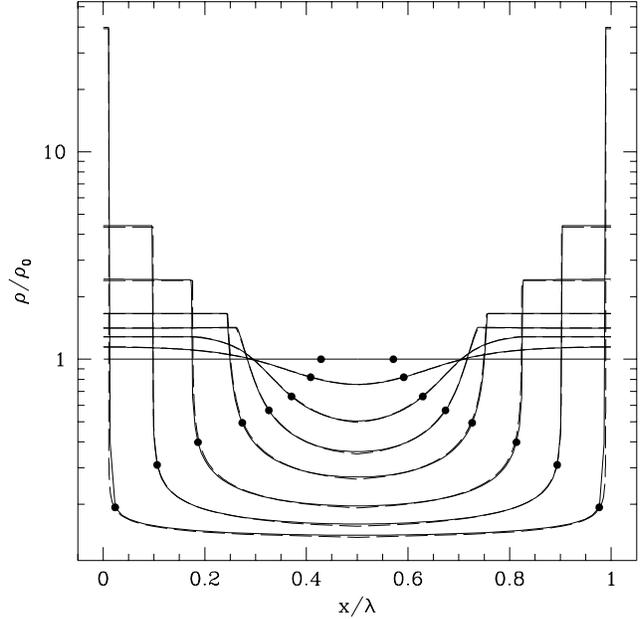}}
 \caption{
The density distributions at subsequent moments
$t/t_0=1.0290$, $1.0300$, $1.0305$, $1.0311$, $1.0321$, $1.0339$ and $1.0560$
for the isobaric S-model (dashed lines)
and hydrodynamical H1-model (solid lines). 
The density distributions of these models are 
almost identical through the evolution.
The full circles
show the locations of the mass coordinate, which will finally divide cold
and hot phases.
}
\end{figure}

\begin{figure}
 \hbox{\psfig{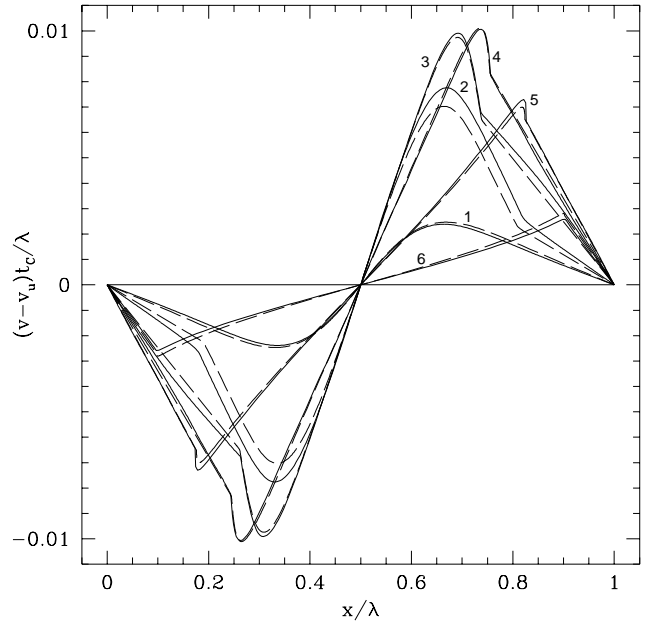}}
 \caption{
The distributions of velocity at subsequent moments
(labeled by the numbers)
$t/t_0=1.0290$, $1.0300$, $1.0305$, $1.0311$, $1.0321$ and $1.0339$
for the isobaric S-model (dashed lines)
and hydrodynamical H1-model (solid lines).
}
\end{figure}

The density and velocity distributions for the H1-model
are presented in Figs~6 and 7 (solid lines) at subsequent moments.
In the same figures the correspondent
distributions for the isobaric S-model (dashed lines) are also shown.
One can see a very good coincidence between these two models.
In Figs~6 and 7 the curves for the H1-model are used with the
time shift $8 t_c$ with respect to the moments for the curves for the
S-model.
This shift follows from  the time delay arising during the linear stage
of evolution of both models (see Appendix~B). The shift is
clearly seen as a small divergence of the $\sigma_T$
curves for the S and H1 models in Fig.~5.

\begin{figure}
 \hbox{\psfig{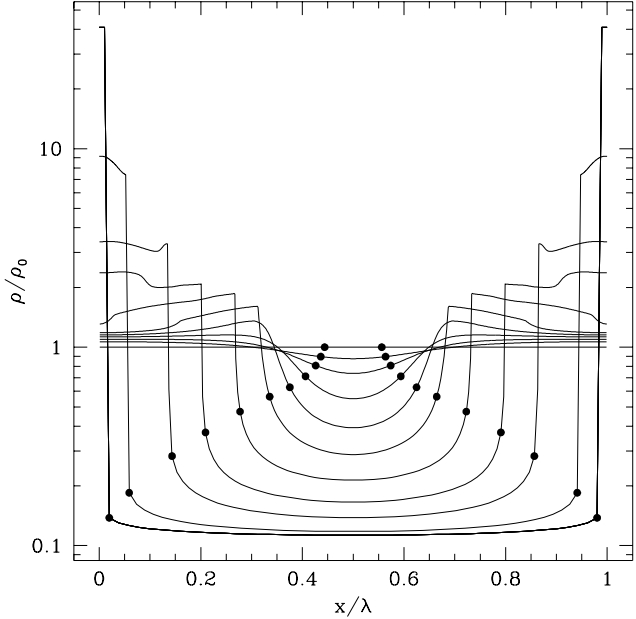}}
 \caption{
As in Fig.~6, but for hydrodynamical H2-model.
The origin of `wave hunches' in the cold and dense phase, 
which are not seen in Fig.~6 for the H1-model,
indicates the breaking of the isobaric condition.
}
\end{figure}

The density distributions of the H2-model
are shown in Fig.~8 at subsequent moments.
The qualitative difference of density distributions in the
H1 and H2 models consists of the origin of `wave hunches' in the latter case,
when the cold phase is compressed by the hot phase.
The wave hunches have a time period of about
$2\pi t_c(\lambda_0/\lambda_P)$.
The origin of wave hunches indicates that the compression zone
of the cold phase at the nonlinear stage is
restricted by the zone of sonic wave
propagation.

The final mass fractions of the cold phase $\eta=0.86$ and $0.885$
for the hydrodynamical H1 and H2-model are shown
in Fig.~4 by a pentagon and square, respectively.
The longer wave H2-model has the larger value of $\eta$.

\section{DISCUSSION AND CONCLUSION}

We have developed a new statistical approach to study the thermal
instability in an optically thin plasma. We investigate
the time evolution of the mass distribution function over temperature
$\varphi(T)$, which characterizes the statistical properties of the
medium. This medium can contain arbitrary spaced three-dimensional
perturbations. We construct the time evolution equation of function
$\varphi(T)$, which describes perturbations of arbitrary amplitude and
is correct under the isobaric condition.
Formally, the obtained evolution equation does not explicitly depend on
the length scale $\lambda$ of perturbations, but
the isobaric condition restricts the characteristic
 length of
perturbations,
$\lambda\la\lambda_P=2\pi c_s t_c$.

The reconstruction of
the space distributions of the density, temperature and velocity
field from the given $\varphi(T)$ is an incorrect procedure
(but we can recover these distributions under the restricted assumptions
of the temperature profile).
However, the statistical approach may be
useful to give us the average parameters
of the inhomogeneous medium, such as the pressure
$P={R\over\mu}\langle\rho\rangle\overline{T}$,
the total
$L=\int \rho^2\Lambda(T)dV=
\langle\rho\rangle M\overline{T}
\overline{\left({\Lambda \over T}\right)}$
and spectral
$L_\epsilon=\int \rho^2\Lambda_\epsilon(T)dV=
\langle\rho\rangle M\overline{T}
\overline{\left({\Lambda_\epsilon\over T}\right)}$
luminosities.

We study the evolution of function
$\varphi(T)$ numerically
in the frame of the slowly expanding cloud scenario (see Section 4).
We follow the thermal instability process in the linear and
nonlinear stages including the formation of a two-phase medium.
In the calculations
we use the cooling-heating function (5.1), which does not allow
the existence of a two-phases equilibrium.
We find that in the case of the expanding medium the
mass fractions of the phases have final nonzero values, contrary to the
case of the stationary (not expanded globally)
medium, where all mass asymptotically tends to the
cold phase.
We also find that the development of a two-phase medium can be suppressed
by a fast expansion, when $t_0/t_c<10^2$.
The final mass fraction of the cold phase is larger
for the slower expanded medium.
Also, the final mass fraction depends on
the amplitude of initial perturbations.
Usually, the larger amplitude produces a larger final mass fraction
of the cold phase.
These results of the calculations are general with respect to the
assumed geometry of perturbations.
In the case of one-dimensional and
periodical
perturbations we compare them with the results of hydrodynamical
calculations. We show that our statistical isobaric approach
correctly describes the thermal instability process
on the length scales up to $\lambda\simeq\lambda_P$.
The hydrodynamical models show a weak dependence on the final mass
fraction on the length scale in case of $\lambda>\lambda_P$.

The presence of a quite strongly tangled magnetic field
of character length scale $\lambda_m\ll\lambda_P$ can suppress
the thermal instability, because the magnetic pressure
will obstruct to the contraction of the cold phase.
In the case of a large scale magnetic field $\lambda_m\ga\lambda_P$
the contraction of cold gas will occur mainly
along magnetic strength lines
forming thin and flat structures
(see also Meerson \& Sasorov 1987 and Sasorov 1988,
where the formation of flattened
plasma condensations was studied).

In the case of a large scale magnetic field $\lambda_m\ga\lambda_P$
the influence of the electron thermal conduction can be important.
Our approach is
valid in the limited time interval, $t_0<t<t_0+\tau+t_\kappa$,
when the hot phase temperature is not too high, $T_h<T_\kappa$.
At temperature $T_h=T_\kappa$ the heating rate (5.2) of the hot phase
is compensated by the cooling effect of the thermal conductivity
$\dot{T}_h=\kappa T_\kappa\mu/R\rho_h\lambda^2$, where
$\kappa=aT^{5/2}$ is the coefficient of thermal conductivity (Spitzer
1962), and
$a=10^{-6}$ in $cgs$ units. In case of $t_\kappa\ll t_0$ the estimation
gives $T_\kappa\simeq 3\cdot 10^5 (\lambda/\lambda_P)^{4/7}K$, and
$t_\kappa\simeq (T_\kappa /T_0)t_c \simeq 30(\lambda/\lambda_P)^{4/7}t_c$.
At short scales
$\lambda\la 10^{-2}\lambda_P$
the thermal conduction suppresses
the thermal instability even in the linear stage (Field 1965).
The perturbations of length $\lambda\simeq\lambda_P$ can grow up to the
maximum temperature $\simeq 3\cdot 10^5 K$. Note that these estimations do
not
depend on the density and heating rate.
The thermal conductivity facilitates the mass exchange between cold and
hot phases. The following evolution of the two phases medium will be
determined by this mass exchange process
(see Aranson et al. 1993; Aharonson et al. 1994).

There is not a full understanding of the problem of dominate scales
of perturbations resulting in the thermal instability process
in the smooth scenario.
It is known that the fastest growth rate has the modes with
$\lambda\la\lambda_P$. But, as it is shown earlier, the electron thermal
conduction can suppress the growth of the shortest wavelength modes.
On the other hand, the longer wavelength modes, $\lambda>\lambda_P$,
have a slower development with an increase of the wavelength.
Thus, there must be a dominate scale, which is determined by the
competition of these factors.
Another important factor, which has an influence on the value of
the dominate scale, is the initial (at the moment $t_0$ of the marginal
stability) spectrum of perturbations. This spectrum depends on the past
history of the medium. We can just note that the linear theory predicts
the dumping of any perturbations at this stage, $t<t_0$,
and that the shorter wavelength modes are dumped faster.
Finally, it seems that the dominate scale will be
located close to $\lambda_P$ for a typical condition
of the optically thin astrophysical plasma.

The possibility of the two-phase medium model for quasars
was first discussed by McCray (1979).
The proposed expanding cloud scenario of thermal instability
can be relevant to the formation of this two-phase medium,
when warm ($T\simeq 10^4 K$) clouds
are formed as the result of thermal instability in
an expanded gas fragment, which is a product of either a star-star
collision (Spitzer \& Saslaw 1966) or star-accretion disc collision
in the $\sim 1 pc$ region around the quasar.
The gas would be heated and ionized
by the UV and X-ray emission of the
quasar.

\section*{Acknowledgments}

We thank for discussion
G. Field, P. Goldreich,  D. Kompaneets,  R. McCray and I. Novikov.
This research was partially supported by the Swedish Natural
Science Research Council and by the Cariplo
Foundation for Scientific Research. We would like to thank
R. Turolla and A. Treves for support during our staying
at the University of Padova and University of Milano,
and R. Nicol and B. H\"ogman for improving the grammar of the text.
The research of AFI was also supported by the grant RFBR 97-02-16975.

\appendix

\section{NUMERICAL METHOD}

\setcounter{equation}{0}

We briefly describe the numerical procedure to solve the equations
(3.12) and (3.15) of the statistical approach. The equations are solved
on the interval  ($T_{\rm min}$, $T_{\rm max}$) of definition
of the distribution function $\varphi(T)$. The interval
is divided on $N$ subintervals by
the grid points $T_{i}$, where $i=0,1,\cdots,N$, and
$T_0=T_{\rm min}$, $T_N=T_{\rm max}$. The evolution of
each point $T_i$ is defined by the equation (3.12).
If the distribution of $\{T_i^n\}$ are known at moment $t^n$,
one can find the new values of $\{T_i^{n+1}\}$ at moment $t^{n+1}$ solving
the system of algebraical equations
\begin{equation}
  {T_i^{n+1}-T_i^n\over\Delta t^n}={1\over 2}\left(F_i^{n+1}+F_i^n\right),
   \;\;\; i=0,1,\cdots,N,  
\end{equation}
where
\[ F_i^k=-{\gamma-1\over\gamma}{\mu\over R}\left\{{\langle\rho\rangle}^k\left(
   (\gamma-1)\overline{\left({\Lambda\over T}\right)}^k
   T_i^k+{\Lambda(T_i^k)\over T_i^k}\overline{T}^k\right)-\right. \]
\[  \qquad\qquad\qquad \left. H^k\left((\gamma-1){T_i^k\over\overline{T}^k}+
   1\right)\right\}-(\gamma-1)\left({\dot{V}\over V}
   \right)^k T_i^k, \]
\[ \overline{T}^k=\sum_{i=0}^{N-1}\int_{T_i^k}^{T_{i+1}^k}
   \varphi^k(T)TdT, \]
\[ \overline{\left({\Lambda\over T}\right)}^k=
   \sum_{i=0}^{N-1}\int_{T_i^k}^{T_{i+1}^k}\varphi^k(T)
   {\Lambda(T)\over T}dT. \]
Here $\Delta t^n=t^{n+1}-t^n$ is the time step. The upper indices
denote the corresponding time. The function $\varphi^k(T)$
is  piecewise-linear distributed through its average
grid values $\varphi_{i+1/2}^k=\Phi_{i+1/2}/(T_{i+1}^k-T_i^k)$,
where
\begin{equation}
   \Phi_{i+1/2}=\int_{T_i}^{T_{i+1}}\varphi(T)dT,
   \qquad i=0,1,\cdots,N-1,
\end{equation}
are conserved quantities,
$  \Phi_{i+1/2}(t)=const $,
as it follows from equation (3.15).
The algebraical equations (A1) are solved by the Newton-Raphson method.
The time step $\Delta t^n$ and the number of iterations are
determined by the assumed accuracy and convergency
of the solution.
Described numerical procedure can not be applied to calculate the
evolution stage of highly inhomogeneous medium, when the time step
becomes very small due to small differences
$T_{i+1}-T_i$.
In this case we modify the numerical algorithm
introducing  a grid redistribution procedure at each time step.
Namely, we choose new values of $\{T_{i}\}$
inside of the interval $(T_{\rm min}, T_{\rm max})$ and remap
$\{\Phi_{i+1/2}\}$ from the old values to the new ones.

\section{LINEAR THEORY IN THE SMOOTH SCENARIO}

\setcounter{equation}{0}

The linear theory of thermal instability (see Weymann, 1960, Field, 1965)
provides us the relations
\begin{equation}
   {P_1\over P_0}=-\alpha{\rho_1\over\rho_0}={\alpha\over 1+\alpha}
   {T_1\over T_0}, 
\end{equation}
where
$ \alpha=\gamma(\lambda n/ 2\pi c_s)^2= \gamma\ell^2x^2$,
between the perturbed pressure $P_1$, density $\rho_1$ and temperature
$T_1$ with a wavelength $\lambda=\ell\lambda_P$, also the velocity
variation $v/c_s=ix\ell\rho_1/\rho_0$, and also the growth rate $n=x/t_c$
of the condensation mode, where $x$ is a positive real root of the equation
\begin{equation}
   \ell^2x^3+\beta\ell^2x^2+x=x_0.
\end{equation}
Other two roots correspond to the dumped wave modes. Here $x_0=n_0 t_c$,
$\beta=\gamma(\gamma-1)$. The scale parameters $\lambda_P$, $t_c$ and
$n_0$ are given by equations (3.1), (3.2) and ($3.17'$), respectively.
We supply here and below the index $0$ to all relevant parameters
(such as $n$, $N$, $\tau$),
which are used in the main text [see formulae ($3.17'$), (4.3)-(4.5)].

In a frame work of the smooth scenario (Section~4), when $\xi=t_c/t_0\ll 1$,
the parameter $x_0(t)$ increases slowly and changes the sign from
negative to positive at $t=t_0$ linearly [equation (4.3)]
\begin{equation}
   x_0(t)=a{\delta t\over t_0}, \qquad \delta t=t-t_0,
\end{equation}
where $a=3b(\gamma-1)$. In this scenario the linear theory
at $\delta t >0$ gives the following:

\vskip 0.1truecm
\noindent
1. The $\ell=0$ case corresponds to the isobaric S-model. We set $\ell=0$
into equations (B1) and (B2), and find $\alpha=0$ and the root $x=x_0$.
The linear stage of the S-model is described by the equations (4.3)--(4.5)
for the $n_0=x_0/t_c$, the integral $N_0(t)$, and the end time $\tau_0$,
respectively. The integral
\begin{equation}
   N_0(t)={x_0^2(t)\over 2a\xi} 
\end{equation}
reaches the $N_0=L_0=\ln (1/A_0)$ level (when the amplitude $A=A_0e^{L_0}=1$)
at the end time
\begin{equation}
   \tau_0=X_0{t_0\over a}={t_0\over a}\sqrt{2aL_0\xi},
\end{equation}
when the end (maximal) growth rate reaches
\begin{equation}
  X_0=x_0(\tau_0)=\sqrt{2aL_0\xi}.
\end{equation}
Note that $x_0\leq X_0\sim \xi^{1/2}\ll 1$, when $\xi\ll 1$.

\noindent
2. The $\ell>0$ cases correspond to the hydrodynamical models
($\ell=1$ for the H1 and $\ell=5$ for the H2-models).
The integral $N_\ell$ has an implicit form now
\begin{equation}
   N_\ell={1\over t_c}\int_0^{\delta t}x\,dt={x^2\over 2a\xi}
   (1+{4\over 3}\beta\ell^2 x+{3\over 2}\ell^2 x^2),
\end{equation}
where we use $dt=t_0 dx_0/a$ [see equation (B3)], $dx_0/dx$ follows
from equation (B2) and $x$ is defined by equation (B2).
The end rate $X_\ell$ can be found from equation
$$ \!\!\,\, 2a\xi L_\ell=
   X_\ell^2(1+{4\over 3}\beta\ell^2 X_\ell+{3\over 2}\ell^2 X_\ell^2),
   \qquad\qquad\qquad \eqno({\rm B7}') $$
which is followed from equation (B7)
where we set the end value $N_\ell=L_\ell=\ln(1/A_{0\ell})$.
Note, that $X_\ell <\sqrt{2aL_\ell\xi}\sim\xi^{1/2}\ll 1$.
The end time $\tau_\ell$ follows from equations (B2) and (B3),
\begin{equation}
   \tau_\ell={t_0\over a}(X_\ell+\beta\ell^2X_\ell^2+\ell^2X_\ell^3),
\end{equation}
where $X_\ell$ is taken from (B7$'$).

\noindent
3. It is useful to compare $N_0$ and $N_\ell$ [equations (B4) and (B7)],
because the difference $\Delta N=N_0(t)-N_\ell(t)$ determines the
suppression of the hydrodynamics modes ($\ell\neq 0$) relative to the
isobaric mode ($\ell=0$). Substituting $x_0$ from equation (B2) to (B4) we
find the difference
\begin{equation}
   \Delta N={\beta\ell^2\over 3a\xi} x^3 g(x),
\end{equation}
where $x$ is determined by (B2) and
\begin{equation}
   g(x)=1+{3x\over 4\beta}[1+2\ell^2(\beta+x)^2].
\end{equation}
It is also useful to compare the end times $\tau_\ell$ and
$\tilde{\tau_0}=\tau_0(L_\ell/L_0)^{1/2}$ [see equation (B5)], where
$\tilde{\tau_0}$ is the end time of $\ell=0$ mode when the initial
amplitude of this mode is equal to $A_{0\ell}$.
The end times difference can be obtained from the form
\begin{equation}
   \tau_\ell^2-\tilde{\tau_0}^2={2\over 3}\left({t_0\over a}\right)^2
   \beta\ell^2 X_\ell^3 g(X_\ell),
\end{equation}
where $X_\ell$ is taken from equation (B7$'$).

\noindent
4. Note that in our case $\xi\ll 1$, when any $x$, $x_0$, $X_0$, $X_\ell$
$\sim\xi^{1/2}\ll 1$, the terms $x^2$ in the brackets
in equations (B7) and (B7$'$),
the terms $x^3$ in equations
(B2) and (B8), and the $x$-term in the bracket in
equation (B10) can be omitted. The equation (B2) is a quadratic now:
$\beta\ell^2 x^2 + x = x_0(t)$ and has a root
\begin{equation}
   x=x(t)=\left(\sqrt{1+4\beta\ell^2x_0(t)}-1\right)/2\beta\ell^2.
\end{equation}
Substituting $x(t)$ in equations (B7) and (B9) we obtain the explicit
time-dependencies of $N_\ell(t)$ and $\Delta N(t)$. We also find
$\alpha(t)$
[see equation (B1)] and can now construct both the
$\sigma_T(t)=\sigma_0e^{N(t)}$ and $\sigma_P(t)=\alpha(t)\sigma_T(t)/
(1+\alpha(t))$ functions on the linear stage, at $0<\delta t<\tau_\ell$.
Graphs of these functions coincide well with the left-hand branches
(at moments $0<\delta t<\tau_\ell$) of the corresponding lines in
Fig.~5.

\noindent
5. In the case $\ell^2x_0\ll 1$ the above written formulae can be
reduced to the simpler forms. From equations (B2, B12) we find
\begin{equation}
   x\simeq x_0-\beta\ell^2x_0^2, \;\;\;\;\; x<x_0\ll{1\over\ell^2}.
\end{equation}
The difference (B9) in the main ($x=x_0$) order is
\begin{equation}
   \Delta N(t)\simeq{\beta\ell^2\over 3a\xi} x_0^3(t). 
\end{equation}
This difference at the end time $\tau_0$ of the $\ell=0$ mode,
when $x_0=X_0=\sqrt{2aL_0\xi}$ [see equation (B6)], is
$$ \!\!\,\,
   \Delta N(\tau_0)\simeq{2\over 3}\beta\ell^2L_0\sqrt{2aL_0\xi}. 
   \qquad\qquad\qquad \eqno({\rm B14}')$$
The end rate $X_\ell$ can be found from equation (B7$'$)
\begin{equation}
   X_\ell\simeq\sqrt{2aL_\ell\xi}
   \left(1-{2\over 3}\beta\ell^2\sqrt{2aL_\ell\xi}\right).
\end{equation}
The end time [equation (B8)] is now
\begin{equation}
   \tau_\ell\simeq{t_0\over a}\sqrt{2aL_\ell\xi}
   \left(1+{1\over 3}\beta\ell^2\sqrt{2aL_\ell\xi}\right),
\end{equation}
and the end times difference [equation (B11)] is
\begin{equation}
   \tau_\ell-\tilde{\tau}_0\simeq{2\over 3}\beta\ell^2L_\ell t_c.
\end{equation}

\noindent
6. For numerical examples we take $b=1$, $\gamma=5/3$, $a=2$, $\beta=10/9$,
$\xi=t_c/t_0=10^{-4}$, $L_0=L_\ell=10$. Results are shown in Table~B1
and can be compared with the results of the statistical theory and
exact hydrodynamical calculations presented in Section~5.

\begin{table}
\caption[ ]{Parameters of the S-model ($\ell=0$),
H1-model ($\ell=1$) and H2-model ($\ell=5$) followed from the linear
theory.} 
\label{symbols}
\begin{tabular}{lccccc}
\hline
$\ell$ & ~~$x(\tau_0)$~~ & ~~$N_\ell(\tau_0)$~~ 
       & ~~$X_\ell$~~ & ~~$\tau_\ell/t_0$~~ & ~~$\alpha(X_\ell)$~~ \\
\hline
$0$ & $0.063$ & $10$ & $0.063$ & $0.0316$ & $0$ \\
$1$ & $0.059$ & $9.56$ & $0.060$ & $0.0323$ & $0.006$ \\
$5$ & $0.033$ & $6.0$ & $0.040$ & $0.042$ & $0.067$ \\
\hline
\end{tabular}
\end{table}


\bsp

\label{lastpage}

\end{document}